\documentclass[preprint,12pt]{elsarticle}
\usepackage{array}
\usepackage{hyperref}
\usepackage{longtable}

\usepackage{amssymb}

\journal{Space Policy}

\begin{document}

\begin{frontmatter}



\title{Space science \& the space economy}


\author[inst1]{Fabrizio Fiore}

\affiliation[inst1]{organization={INAF Osservatorio Astronomico di Trieste},
            addressline={via G.B. Tiepolo 11}, 
            city={Trieste},
            postcode={I34143}, 
            country={Italy}}

\author[inst2]{Martin Elvis}

\affiliation[inst2]{organization={Center for Astrophysics | Harvard \& Smithsonian},
            addressline={60 Garden St.}, 
            city={Cambridge},
            postcode={02138}, 
            state={MA},
            country={USA}}

\begin{abstract}
Will it be possible in the future to realize large, complex space missions dedicated to basic science like HST, Chandra and JWST? Or will their cost be too great? Today's space scene is completely different from that of even five years ago, and certainly from that of the time when HST, Chandra and JWST were conceived and built. Space-related investments have grown exponentially in recent years, with a monetary investment exceeding half a trillion dollars per year since 2023. This boom is greatly aided by the rise of the so-called 'new space' economy driven by private commercial funding, which for the first time last year surpassed public investments in space. The establishment of a market logic to space activities results in more competition and a resulting dramatic cost and schedule reduction.  Can space science take advantage of the benefits of the new space economy to reduce cost and development time and at the same time succeed in producing powerful missions in basic science? The prospects for Europe and the United States are considered here. We argue that this goal would be achievable if the scientific community could take advantage of the three pillars underlying the innovation of the new space economy: (1) technology innovation proceeding through both incremental innovation and disruptive innovation, (2) business innovation, through vertical integration, scale production, and service-oriented business model, and (3) cultural innovation, through openness to risk and iterative development.
\end{abstract}




\end{frontmatter}


\section{Basic science from space}
\label{sec:one}
Astrophysics is a field that has been in permanent revolution for nearly 200 years, since the invention of spectroscopy in 1859 let us learn the composition of the Sun and stars. In the past 25 years alone, at least four major, unanticipated, astrophysics topics have emerged and conquered the scientific 'market'.
\begin{itemize}
\item gravitational wave astrophysics and multimessenger astrophysics
\item exoplanets
\item black holes - galaxy scaling relations - black hole direct imaging
\item the accelerated expansion of the Universe.
\end{itemize}
The first was still beyond the horizon when the James Webb Space Telescope (JWST, at the time called Next Generation Space Telescope, NGST) was recommended at the end of the 1990s as the highest priority in the 2000 Decadal Survey. The other three were in their infancy, and the first discoveries were just announced.  They were not among the main science drivers for NGST. Instead, in the Decadal report 'Astronomy and Astrophysics in the New Millennium' 
\cite{decadal2000} one reads that NGST {\it '..is designed to detect light from the first stars and to trace the evolution of galaxies from their formation to the present. It will revolutionize the understanding of how stars and planets form in our galaxy today.. thanks to  Kuiper Belt objects (KBOs) in our solar system, the formation of stars and planets in our galaxy, and the dust emission from galaxies out to redshifts of 3'}. JWST is doing superb work on the earliest galaxies. Yet JWST is doing a great job on all four of the new themes: up to 40\% of JWST time is dedicated to exoplanet studies alone. JWST can be this flexible because it was designed to be a multipurpose observatory that pushes the sensitivity for both photometry and spectroscopy by orders of magnitude with respect to previous or contemporary facilities, opening up new 'discovery space.' This powerful capability did not come for free. The complexity, cost and development time of JWST are large, even huge, with more than 300 single-point failure modes \cite{Kalia2023}, a cost of about 9B\$ and 23 years elapsed from the publication of the 2000 decadal survey to launch. This cost should be compared with the predicted cost to completion for NGST in 'Astronomy and Astrophysics in the New Millennium', given as 1B\$. This almost 10-fold cost growth of NGST cannibalized other NASA Astrophysics programs, notably the Beyond Einstein program that spanned a wide range of other wavelengths \cite{NASABE}. Concentration on NGST introduced an enormous risk to space-based astrophysics by putting all its eggs in one basket: A failure of JWST would have meant a decades-long setback for all observational astrophysics.  

Twenty years after the 2000 decadal report recommendation of NGST, a new National Academies  decadal survey, Astro 2020,  recommended a “large (~6 m aperture) infrared/optical/ultraviolet (IR/O/UV) space telescope” (now called the Habitable Worlds Observatory, HWO) as its highest priority, next generation telescope \cite{decadal2020}, with a deployment date scheduled for sometime in the 2040s and at an estimated cost of \$11 B (a timescale and cost pattern similar to that of the as-flown JWST). As the name suggests, the main scientific goal is imaging and spectroscopy of Earth-like planets using extremely high suppression ($>10^{10}$) coronagraphs. 

By the time of HWO launch, somewhere in the next 20-25 years, new scientific themes are likely to emerge, and exoplanets, which today are probably the primary astrophysical topic, may fade somewhat in relative importance. Will the discovery space of HWO be large enough to accommodate other breakthrough discoveries, as for JWST? The Astro2020 wording does include a recommendation that HWO span {\it'… a wide range of astrophysics'}. In practice, this recommendation seems to be taken as a requirement to have a good UV response, preferably shortwards of Lyman-alpha\footnote{\url{https://ntrs.nasa.gov/citations/20230014335}}, which is technically demanding and pulls against the high-suppression level coronagraph. Extremely tight and sometimes in tension requirements mean long and expensive work to address them. Today, a good UV response is not a driver of exoplanet or cosmology research. By 2045 will UV be the key to making progress in the, as yet unknown, new fields?

Even more worrisome is that it is far from certain that NASA, ESA, or any other space agency will be willing or able to embark on a new JWST-scale project (i.e., \$10B, $>20$ year development time). In April 2024 NASA paused the original JPL-led Mars Sample Return project\footnote{\url{https://www.nasa.gov/news-release/nasa-sets-path-to-return-mars-samples-seeks-innovative-designs/}}, which was in the ballpark cost and development time range as JWST. NASA issued a call to industry for alternative approaches and in June 2024 selected seven companies to conduct studies of 'out of the box' concepts that can deliver samples faster and cheaper than the current plan\footnote{\url{https://www.nasa.gov/news-release/nasa-exploring-alternative-mars-sample-return-methods/}}. This rethought mission may become another example in which NASA buys a service for science-related or exploration-related missions, as NASA now does in other areas, for example, transportation to and from the International Space Station and the Commercial Lunar Payload Service. Another example of the NASA change in attitude is the end of the Moon VIPER rover project\footnote{\url{https://www.nasa.gov/news-release/nasa-ends-viper-project-continues-moon-exploration/}} 
due to increases in costs and delays.All this shows that we may have already hit the funding wall \cite{Elvis2016}.  An additional cause for concern to astrophysics is that NASA's major effort today and for the next few years will certainly be the Artemis project, returning humans to the lunar surface in a long-term program. How can space astrophysics thrive in these circumstances? Can projects like HWO and Artemis coexist? An even greater element of uncertainty is to predict at this time what the attitude of NASA and the new U.S. government in general toward basic science missions might be in the next years. The first signals are definitely not encouraging, given the proposed cuts to the NASA 2026 budget (a two-third cut to NASA astrophysics with respect to 2025, a third cut to planetary science, and a half cut to Earth science\footnote{\url{https://www.nasa.gov/wp-content/uploads/2025/05/fy-2026-budget-technical-supplement-002.pdf}}.

An alternative to expensive multipurpose observatories is to have multiple dedicated missions. ESA’s Gaia mission, for example, opened up new discovery space by measuring the positions and motions of a billion stars some 1000 times better than was possible before. Gaia is becoming the most productive telescope in terms of the number of papers produced from its data, about 2000 papers published per year in refereed journals\footnote{\url{https://www.cosmos.esa.int/web/gaia/peer-reviewed-journals}}, compared with roughly 1000 papers per year based on HST data\footnote{\url{https://science.nasa.gov/mission/hubble/overview/hubble-by-the-numbers/}}. The ESA Euclid and NASA Nancy Grace Roman telescopes are other examples. The total cost of Gaia and Euclid (ESA plus member state contributions) is in the 1-1.5B\$ each, while that of the Roman telescope is close to 5 B\$
(adding a coronagraph to Roman increased its cost by \$ 350-500M\footnote{\url{https://roman.gsfc.nasa.gov/science/rsig/2021/Roman\_Requirements\_20201105.pdf}}).

Their development time is in the 10-15 years range. That is about 2-10 times cheaper and 2 times faster than JWST and HWO. Although this is an improvement, even at these costs, it is not clear that space agencies can afford more than one per decade of these missions. Such a pace creates the risk that many emerging scientific topics, and so sectors of the astrophysical community, are left unserved at the margin.

Unlike ESA, NASA has a strong Explorer program\footnote{\url{https://explorers.gsfc.nasa.gov/}} that provides affordable and frequent flight opportunities for smaller science missions \cite{Elvis2009}. This ensures a rapid response to changing science and technology to enable cutting-edge science at moderate cost. During the long JWST development time, the Explorer program has contributed to maintaining a reasonable level of diversity in the astrophysical ecosystem, avoiding the drying up of entire scientific communities. In Europe, the lack of an Explorer-like program has challenged the high-energy, mid/far-infrared and millimeter communities, which were unable to benefit from replacements for the highly successful XMM-Newton, Herschel and Planck X-ray, far-infrared and CMB-millimeter missions. Of course, Explorer missions are somewhat limited in scope, being dedicated to well-focused scientific goals and in some cases representing case studies and in-orbit demonstrations for new techniques. As an example, the latest SMEX astrophysics mission is IXPE, dedicated to X-ray polarimetry, certainly a niche science. Although IXPE is certainly a highly successful mission (the team was awarded the 2024 Bruno Rossi Prize), its limited collecting area allows sensitive observations of only a few dozen bright sources. IXPE has shown the high potential of X-ray polarimetry, but to fully harvest its fruit, a much larger X-ray polarimetric telescope is needed. The current cost cap of Explorer missions is around \$120M - \$130M for the SMEX missions and about \$300M for MIDEX missions\footnote{\$300M in NASA Fiscal Year (FY) 2022 dollars: \url{https://nspires.nasaprs.com/external/viewrepositorydocument/cmdocumentid=841540/solicitationId=\%7B0F29B64B-AFF1-ADAD-7302-75D7AEF2C696\%7D/viewSolicitationDocument=1/MIDEX_AO_Amend\%201.pdf}}. From 2000 seven SMEX and five MIDEX astrophysics missions were launched\footnote{\url{https://en.wikipedia.org/wiki/Explorers_Program\#Small_Explorers_(SMEX)}}, that is about a mission every two years. The typical development time of Explorer missions is 4-5 years, and the entry-level TRL\footnote{TRL = technical readiness level: \url{https://en.wikipedia.org/wiki/Technology_readiness_level}} is 5-6. 

Rapid cadence, short development time, and high TRL entry level make Explorer missions a good starting point, together with the cubesat approach, for a new generation of ambitious yet affordable space-science missions (Section 4).

\section{The new space economy}
\label{sec:two}

At the same time that JWST was launched and HWO was being conceived, we witnessed the booming of the so-called Space Economy. The total amount of money invested in the Space Economy in 2023 was \$570B according to The Space Report of the Space Foundation\footnote{\url{https://www.spacefoundation.org/2024/07/18/the-space-report-2024-q2/}}, with a growth of 7.4\% with respect to 2022. According to the Novaspace Space Report 2024 the total size of the Space Economy in 2024 was \$596B \footnote{\url{https://nova.space/hub/product/space-economy-report/}}, with a further growth of 4.5\% compared to the previous year. Recent years have seen private funding overtake public funding. Growth may well accelerate in the next few years, reaching the threshold of \$1000B at the end of this decade and up to 1.8 trillion dollars by 2035\footnote{\url{https://www3.weforum.org/docs/WEF_Space_2024.pdf}}, dwarfing any conceivable increase in government space agency budgets (except maybe military). Many of these start-ups go well beyond providing launch into on-orbit applications such as Internet communications and many forms of Earth observation. Satellite numbers have grown from less than 200/year in 2000, to almost 3000/year in 2023\footnote{\url{https://www.planet4589.org/space/stats/pay.html}
}.

An important indicator of the strategic role that space is gaining in private businesses is the amount of money invested by venture capital firms in space startups. Between 2006 and 2014 it was of the order of  0.1B\$/yr, from 2019 to today it was on average 8B\$/yr\footnote{Start-up space 2023, Bryce Tech, \url{https://brycetech.com/reports}}, an almost two orders of magnitude growth. The number of investors is today of the order of 500, with a constant fraction of first time space investors. On the other hand, private equity firms are less active in the space business, investing less than 10\% than venture capital firms. The result, especially in Europe, is that while seed money is relatively easily available for startups, space companies sometimes struggle to find investors for later stage A, B, and C rounds. In any case, venture capital investment is willing to take higher risks, for occasional enormous rewards. If so many companies are today investing in new space startups, they must see opportunities to make huge profits in the coming years \cite{Anderson2023}.

The complete entry of space activities into the market logic results in increased competition, cost reduction, and introduction of timing as the main drivers. (In commerce, time is literally money). Using a time-saving approach produces both cost savings and the ability to go to market quickly. 

Can the scientific community take advantage of the new space economy to produce ambitious new space missions dedicated to basic science at a much lower cost and faster timeline? One of the authors, M. Elvis, already discussed what the commercial sector can do for us scientists \citep{Elvis2016}. Along the same lines, \cite{Metzger2016} and \cite{Crawford2016} discuss how space science can benefit from the development of a robust space infrastructure. Here we review all these issues in light of the booming of the new space economy and discuss further questions, such as what the scientific community should do to really exploit the opportunities provided by the commercial space economy. Before addressing these questions, we need to better understand the differences between the old space economy and the new. Table 1 provides a useful simplification as a first approach.

\begin{longtable}[H]{ | m{6.4cm} | m{6.4cm}| }
\caption{Old vs New space economy contrasts. The numbers quoted are from the World Economic Forum 2024 report $^*$}\\ \hline
{\bf Old space economy}  & {\bf New space economy} \\ \hline
Public funding, largely dependent on politics. Contracts may be awarded by geography as well as merit & Private funding. Most private investors seek economic profit. However, there are also foundations and non-profit organizations that may be willing to invest in space science missions for other reasons. See \cite{MacDonald2017} for examples of large scale philanthropic support for astronomy \\  \hline
Expenditure is decided by Governments and limited by appropriation budget
& Expenditure is decided by market and by the company's policies  \\ \hline
Strategic returns  & Commercial return \\ \hline
Indirect uptake by households/consumers & Direct uptake by households/consumers \\ \hline
2023 expenditure: ~300 B\$ & 2023 expenditure: ~300B\$ \\ \hline
Examples of current major activity/business in order of largest to smallest share: not publicly funded communications, 40\%; defence, 20\%; positioning, navigation, and timing (PNT), 14\%; infrastructure, 6.6\%; exploration, 4.5\%; Earth observation, 4.2\%; basic science and space observation, 3.3\% &
Example of current major business enabled by activities, in order of largest to smallest share: PNT, 65\% (a large fraction is for chips in smartphones, financial transactions use PNT timing); communications, 22\%; defense, 9.3\% \\ \hline
Future: The public-funded space economy will more than double by 2035.  Expenditure on defense may triple. Other areas of great increase: PNT;  infrastructures. Other areas of public investment for technology development and demonstration: manufacturing in space, including pharmaceutics, biotechnology and semiconductors; space mining; space solar production; space travel; &
In the past decades activity/business started in the old space-economy framework using public money and then progressed to the new space economy. By 2035 the new space economy may more than triple. Areas of large increases will probably be PNT, which alone may consist of $>$75\% of all business, for example last mile deliveries and point-to-point transportation; communications, Earth observation. Other areas can reach maturity and enter the commercial sector.  Among them: manufacturing in space, space travel, space mining, space solar production.  \\ \hline
Science data is public (largely); military not & Data is IP \\ \hline
\caption{$^*$ \url{https://www3.weforum.org/docs/WEF_Space_2024.pdf}}
\end{longtable}

The business in space is enabled by resources. Until now, the main space resource exploited is location: the special position in Low Earth Orbit (LEO), $>$400-600 km from the Earth’s surface, Medium Earth orbit (MEO) and Geostationary orbit (GEO) enabled some of the major businesses currently on going: PNT, communications, Earth observation. The "location" resource will maintain and even increase its strategic importance in the future. Other resources associated with location are microgravity and special environments. Today we are still in the research and development phase for production of goods in microgravity, but the situation can quickly change in the next few years with the advent of private space stations. The other commonly used resource in space is solar light, to produce power for satellites and energy for thermal control systems. In the future space solar power production could also be used for Moon and Earth applications, by beaming power to the ground using infrared and submillimeter radiation. 

Another important space resource will be materials \citep{Elvis2021, Crawford2023}: water, oxygen, hydrogen, construction materials including iron, titanium, and rock, and precious metals, including the Platinum group elements (PGE). Rare Earth elements (REE) are also common on the Moon in the KREEP terrane \citep{McLeod2017}. REE concentration in the Apollo samples is lower than that of typical REE ores on Earth, but of course the number of samples is limited, and therefore a more detailed prospecting for such elements in promising areas (e.g. Oceanus Procellarum) is desirable. REE were subject to large price volatility in the past decades, in particular when China put strict export quotas in 2009-2010 \cite{Haque2014,Barakos2016}. Helium 3 concentration on the lunar surface is higher than in most Earth sites, and this element is becoming strategic, for example, to support quantum computers. It is very likely that the first use of materials extracted from the Moon will be in situ, to support astronaut activity (water, oxygen, construction materials) or to produce rocket propellant (hydrogen, oxygen). This will be the first driver of the cis-lunar economy. Only at a later stage it would be conceivable to mine for precious metals or strategic elements that would be brought back to Earth. 

The final important resource commonly used in space, but not often acknowledged, is "knowledge". Knowledge is crucial to enable or make more efficient harvesting of other resources and to protect space systems. Navigation in low Earth orbits requires detailed knowledge of the Earth’s gravitational field; likewise for the Moon. Navigation near small bodies requires accurate knowledge of their gravitational field and response to solar illumination. Without a deep understanding of the physics of the solar wind and without real-time data, it is impossible to forecast solar storms that can easily damage space infrastructures. And, it is impossible to mine for resources the Moon or Mars or an asteroid if it is not known with precision what is present on the surface or interiors of these bodies through prospecting. These are just a few examples, but it is clear that without long and accurate observations, big data analysis, and the development of accurate models it is impossible to imagine any activity in space. Knowledge therefore has its own very concrete value as intellectual property (IP)  in the new space economy. This is a major change from the open data policies of most government space agencies.

\subsection{Innovation paradigms}

Table 2 compares the innovation paradigms that underlie the old and new space economy. 

\begin{longtable}[H]{ | m{6.4cm} | m{6.4cm}| }
\caption{Strategic versus Commercial Paradigms} \\
\hline
\bf{Old space economy: strategic paradigm}    & \bf{New space economy: commercial paradigm} \\ 
\hline
Geopolitical influence: low risk but limited initiative. There is no incentive to reduce costs. In fact, this is an incentive to keep space expensive so that only a few can afford it. Perpetuates the "space is hard" attitude &
Make money: free market but private capital needed, as well as government support \\
\hline
Military advantage: no need for private capital &
Quality of life, wellness is proportional to innovation, but innovation has hardly decreased inequalities, both vertical and horizontal\\
\hline
Technological development is pushed: start from  TRL 0 but barriers at TRL 6 &
Technological development is pulled,  start at TRL$>$5-6 \\
\hline
Need new invention & Allow new invention\\
\hline
High cost &  Low cost \\
\hline
Single missions &  Constellations \\
\hline
Artisanal production & Scale production \\
\hline
Distributed procurement & Vertical integration \\
\hline
Risk averse, waterfall project management & Risk openness, Agile iterative design project management \\
\hline
Timing is not usually a driver, and fast delivery is not mandatory. In contrast, in many cases, contracts were made on a cost reimbursement basis, so schedule delays actually increased a company's bottom line &
Timing is a driver, delays must be contained to increase return of investment. Fast delivery of products and services is mandatory \\
\hline
\end{longtable}

There are three pillars that are the foundations of the new space: 
\begin{enumerate}
    \item 
Innovation in technology, proceeding through both incremental innovation and disruptive innovation. Disruptive innovation starts with low performance and high cost, but quickly overtakes mainstream technologies. The disruptors then engage in incremental innovation, beginning a new cycle. 
\item 
Business innovation, defined as vertical integration, scale production and service oriented business model. 
\item 
Cultural innovation. Startups must have a bold risk culture, which manifests itself in risk openness project management models, for example the Agile methodology, which breaks projects into several 'sprints' and delivers a Minimum Viable Product (MVP) at the end of each sprint. 
\end{enumerate}

In the next sections, we will elaborate on the main sources of costs of space missions.

\subsection{Costs: launch}

\begin{figure}[ht]
\includegraphics[width=10cm,angle=-90]{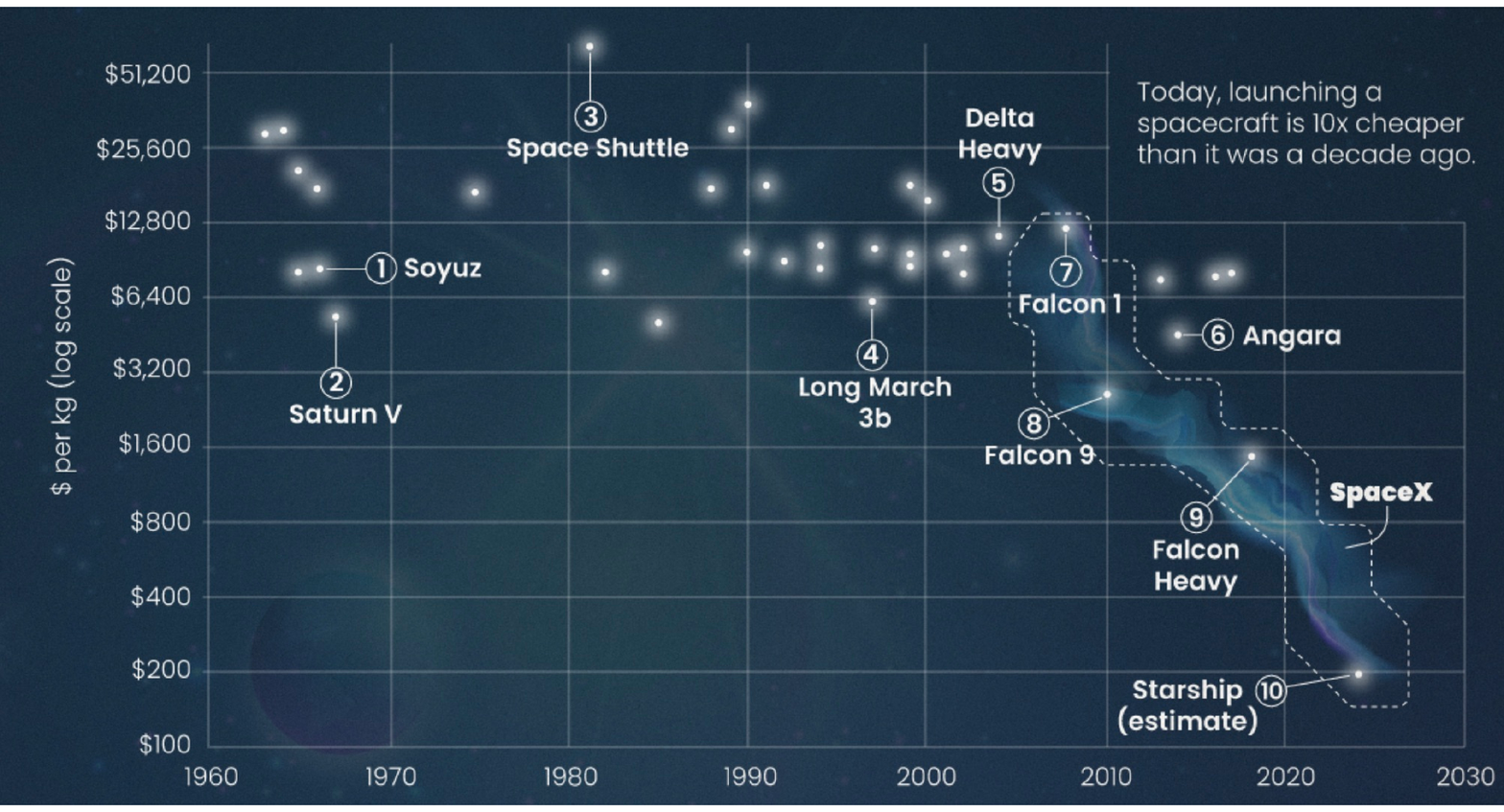}
\caption{Cost/kilogram to LEO versus year}
\label{launchcost}
\end{figure}

Figure \ref{launchcost}, adapted from Visual Capitalist\footnote{\url{https://www.visualcapitalist.com/the-cost-of-space-flight/}}, shows the cost per kg to orbit as a function of time. Due to reusability, the SpaceX Falcon 9 (F9) and Falcon Heavy (FH) launchers have lowered the cost to orbit offered to third parties by a factor of 3-6 with respect to other Western, Russian and Chinese launchers. F9 capacity to low Earth orbit (LEO) is about 20 tons, at an advertised cost of about 62M\$, that is about \$3000/kg.

Today, F9 has an effective monopoly on the launch market. Forthcoming competition with Blue Origin New Glenn may result in a lowered cost for third parties. Table 3 provides examples of launch costs for recent scientific missions. None approach the \$1-3k/kg that is achievable with F9, mainly because they were unable to make use of the full F9 capacity on dedicated launches, especially for missions not designed for an F9 or FH launch. Other causes of high-billed costs include requests for orbits more demanding than the LEO and absence of competition.

\begin{longtable}[H]{ | m{2cm} | m{1.5cm}|  m{2.5cm} | m{2cm} | m{1.25cm} | m{1.25cm} |}
\caption{Mission launch cost} \\
\hline
Mission & Mass kg & Orbit & Launcher & Cost M\$ & Cost k\$/kg \\
\hline
IXPE & 330 & LEO equatorial & F9 & 50.3 & 152 \\
\hline
Euclid & 1500 & L2 & F9 & $\sim70$ & 47 \\
\hline
Psyche & 2750 & Interplanetary & FH & $\sim$131 & 48  \\
\hline
Europa clipper & 6000 & Interplanetary & FH & 178 & 30 \\
\hline
TESS & 362 & HEO & F9 & 87 & 240 \\ 
\hline
ICON & 288 & LEO & Pegasus XL (F44) & 60.3 & 195 \\
\hline
COSI & 365 & LEO equatorial & F9 & 69 & 189 \\
\hline
JWST & 6500 & L2 & Ariane 5 & 189$^a$ & 29  \\
\hline
Juice & 2420 & Interplanetary & Ariane 5 & 162$^a$ & 67 \\
\hline
Solar Orbiter & 1800 & Interplanetary & Atlas V 411 & 172.7 & 96 \\
\hline
Bepi Colombo & 4100 & Interplanetary & Ariane 5 & 162$^a$ & 40 \\
\hline
Gaia & 2030 & L2 & Soyuz ST-B/Fregat-MT & 86.4$^a$ & 42 \\
\hline
\caption{$^a$1EUR=1.08\$}
\end{longtable}

The cost of launching cubesats to Low Earth Orbits (LEO) is obviously much lower because of the very small mass of these satellites. Currently, the SpaceX Transporter F9 ridesharing program offers launches at about \$6.000/kg, although these costs can easily be multiplied by a few when buying launches indirectly from launch providers\footnote{\url{https://www.nanosats.eu/tables\#launch-providers}}. Launching from the International Space Station is more expensive by a factor of a few\footnote{\url{https://www.satcatalog.com/insights/cubesat-launch-costs/\#:~:text=The\%20cost\%20for\%20CubeSat\%20deployment,\%24270\%2C000\%2C\%20and\%20a\%206U\%20\%24540\%2C000}}. The launch cost of the micro-launcher Electron is even higher, reaching about \$35.000-40.000/kg\footnote{\url{https://www.newspace.im/launchers/rocket-lab}}.

The launch cost of Figure \ref{launchcost} has been reached for SpaceX Starlink satellites that have been estimated at 0.7-1.4 M\$ per Starlink 2 mini satellite, with a cost per kg as low as \$1000-\$2000/kg. This should be regarded as the cost of launch for optimized payloads, net of profit. SpaceX claims that the coming on-line of Starship in the next few years will lower the cost by another factor of 10 with respect to F9. The Starship revolution will not be limited to cost to orbit. Starship performance will make it unnecessary to spend time and resources to minimize mass and size \cite{Anderson2023, Elvis2023, Heldmann2021, Douglas2023}, which will change not only the way satellites are built, but, more importantly, the way missions are architected. In fact, this paradigm shift is already underway. As an example, for SpaceX Falcon9 Transporter missions, one can decide which interface plate to buy and then fit on that plate a satellite of mass and volume up to the maximum allowed by that plate, at a fixed cost\footnote{\url{https://rideshare.spacex.com/search}}.

\subsection{Costs: satellite production}

The second major cost of a space mission is the cost related to the satellite procurement, integration, and test. Table 4 reports the cost of several scientific satellites, including payloads, bridging the typical cost of large flagship missions and small explorer missions (1.4-0.42M\$/kg). For comparison, we also report the costs of standard cubesats using COTS components, advanced cubesat, cubesats for commercial constellations, and the SpaceX Starlink satellites. 

\begin{longtable}[H]{ | m{2.5cm} | m{2cm}|  m{2cm} | m{2cm} | m{2cm} | }
\caption{Satellite Costs, including bus and payloads} \\
\hline 
Satellite & Multiplicity & Mass kg & Cost M\$ & Cost M\$/kg \\ \hline
JWST & 1 & 6500 & 9000 & 1.4 \\ \hline
Europa clipper & 1 & 6000 & 5200 & 0.9\\ \hline
Euclid & 1 & 2000 & 1510$^a$ & 0.76a \\ \hline
IXPE (SMEX) & 1 & 330 & 138 & 0.42 \\ \hline
HERMES-PF 3U & 6 & 6 & 0.86$^a$ & 0.144a \\ \hline
Standard 3U CubeSat &  & 5 & 0.1-0.4 & 0.02-0.08 \\ \hline
 Planet Lab 3U & 100 & 5.2 & ~0.1-0.4$^*$ & 0.02-0.08 \\ \hline
Skysat & 10 & 110 & ~3.5* & ~0.032 \\ \hline
Starlink1 & 1000 & 260 & ~0.25$^*$ & ~0.001 \\ \hline
Starlink 2 mini & 1000 & 730 & ~0.8$^*$ & ~0.0011 \\ \hline
Starlink 2 & 1000 & 1500 & ~1.2$^*$ & ~0.0008 \\ \hline
\caption{$^*$net cost of the hardware/software, not including profit for the manufacturer; $^a$1EUR=1.08\$}
\end{longtable}

The cost/kg of standard cubesats is 5-20 times lower than the cost/kg of Explorer missions. Of course, standard cubesat are very simple and limited. A better comparison can be made with advanced cubesats, designed specifically for a scientific mission. One good example is HERMES Pathfinder (HERMES-PF), a European constellation of 6 3U cubesat equipped with miniaturized but advanced X-ray and gamma-ray detectors to observe cosmic explosions such as Gamma Ray Bursts and the electromagnetic counterparts of gravitational wave events \cite{Fiore2020,Fiore2021,Fiore2025}. The net cost of each HERMES-PF 3U CubeSat is $\sim$\$860.000, including payload (P/L). The spacecraft (S/C) uses selected COTS components, procured through contracts with companies including non-recurring engineering phases to tailor the hardware to the specific application.  The P/L and the S/C do not use radhard Electrical, Electronic and Electro-mechanical (EEE) components. HERMES-PF S/C cost/kg is about 145.000 \$/kg (the cost/kg of the P/L is similar),  or 2-7 times higher than the standard 3U cubesat but only one-third of that of Explorer missions. The cost of cubesats and mini/midi-sats for commercial constellations is similar or lower to that of standard cubesats, at a level of 20.000-30.000\$/kg \footnote{\url{https://sustainability.e-shape.eu/economics-behind-satellite-megaconstellations/https://www.bloomberg.com/news/features/2017-06-29/the-tiny-satellites-ushering-in-the-new-space-revolution}}

The cost of SpaceX Starlink satellites (bus and P/L included) is lower than that of cubesats by more than another order of magnitude, reaching an incredibly low value of about 1000\$/kg\footnote{\url{https://spacenews.com/starlink-soars-spacexs-satellite-internet-surprises-analysts-with-6-6-billion-revenue-projection/}}. Interestingly, this cost/kg is about constant for all Starlink satellites despite a weight difference of a factor of 6. This is 20-30 times lower than the cost of a standard cubesat and that of advanced cubesat and mini-sat for Earth observation, more than 100 times lower than the cost of advanced cubesats for science applications, and more than 400 times lower than the cost of Explorer missions.

\section{Public Funding Pros and Cons for Large Missions}

So far science missions in space have been paid almost entirely by public funding (governments), and thus they belong to the 'old space' economy. Table 5 summarizes the pros and cons of this situation.

\begin{longtable}[H]{ | m{6.4cm} | m{6.4cm}| } 
\caption{Pros \& Cons of public funding for science in space}\\
\hline
Pros & Cons\\ \hline
Basic science missions are possible, the portfolio is not limited to missions in applied science. Missions can address problems that may not have direct impact on today or tomorrow quality of life/wellness (they may have impact in 50-100 years, or may not have impact at all) &
Single mission, sometimes high risk / high gain (e.g. JWST). Large chunks of community disbanded after lack of new missions in the field, producing loss of competence/skills;
generation-long development time (20-30 years in some cases.) 
Opportunity cost: What missions were not made to pay for JWST-class missions? \\ \hline
Funding is not or is not directly related to quarterly profits and is therefore more stable than in private sector &
Intrinsically limited funding based on taxation, see figure\ref{budget}.
 \\ \hline
Pushes technological development &
Artisanal production. Difficult to replicate (e.g. Chandra) at very high cost \\ \hline
Produces spin-offs &
Limited incentives for spin-off leading to reduced efficiency in innovation with respect to the private sector.  \\ \hline
Possible to exploit synergies with other strategic programs (military for example), see IR detectors, Swift agility, Roman mirror, HST & \\ \hline
\end{longtable}

\begin{figure}[ht]
\includegraphics[width=6.8cm]{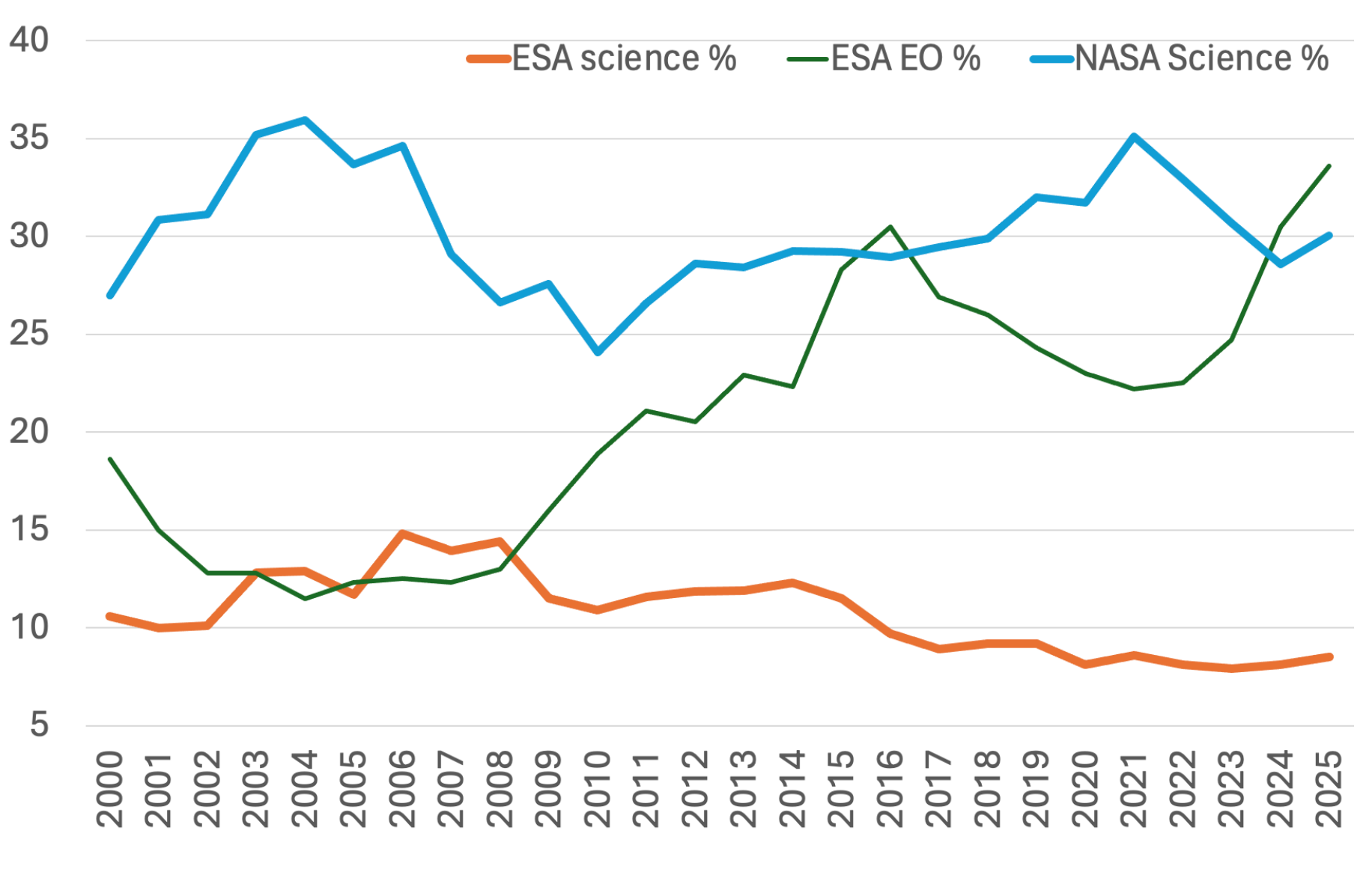}
\includegraphics[width=6.8cm]{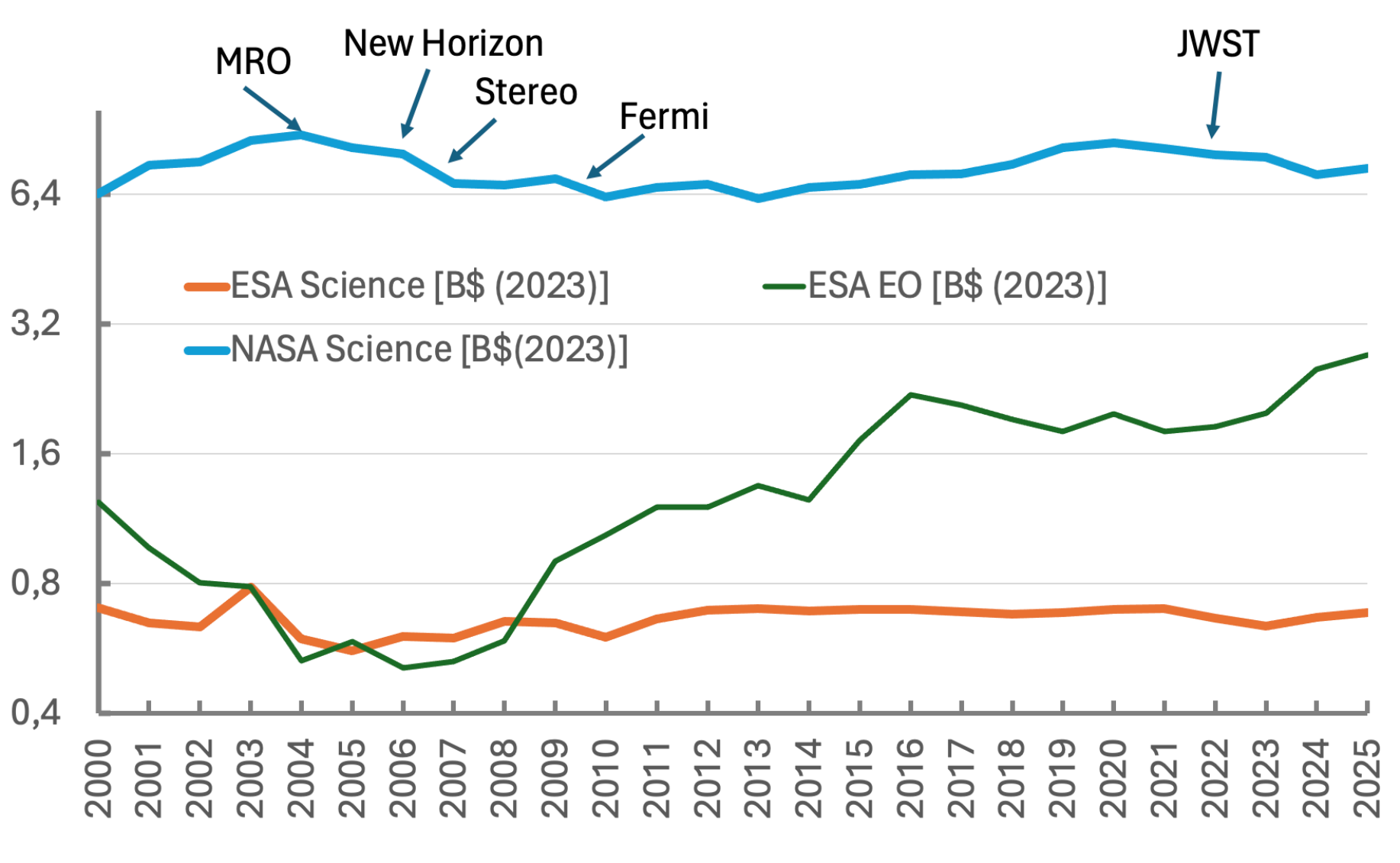}
\caption{[left panel] The percentage of the total NASA and ESA budget that went in science and Earth observation mission from 2000 to 2025. NASA science budget include Earth observation science. ESA Earth observation includes both science and applications. [Right panel] The NASA and ESA budget for science and Earth observation in \$(2023), therefore corrected for inflation.}
\label{budget}
\end{figure}

Figure \ref{budget} shows the evolution of NASA and ESA budgets since 2000. NASA's budget for science, including solar system exploration, Mars exploration, astrophysics, and Earth observation, varied between 25\% and 35\% of the total between 2000 and 2025. The ESA budget for science decreased from about 15\% of the total in 2006-2008 to 8.5\% in 2020-2025. This does not include Earth observation, which includes in the ESA system both science and application (the Copernicus program). Figure \ref{budget} also shows NASA and ESA science budgets in absolute values, corrected for inflation to 2023. Note the peaks in the NASA budget around 2003-2006 and 2018-2022. The second is mainly due to spending on JWST. The first is due to spending on several smaller missions, including Messenger, MRO, New Horizon, Dawn, Stereo and Fermi, witnessing a period of great diversification in NASA science missions. ESA science budget remained remarkably constant around a value of 600-700 M\$(2023) per year for about 25 years (the 2003 budget was increased by about 100MEUR to take stock of the grounding of Ariane 5, which postponed the launch of Rosetta). ESA budget between 1986, following the agreement of the Horizon 2000 plan, and 2013 is presented by \cite{Southwood2014}. The budget peaked in 1996 at about M\$(2023), when a negative gradient started because at the 1995 ESA Council Ministerial meeting a flat cash was imposed. The Horizon 2000 plan established Europe as a major player in space science and drove technological innovation in Europe. It was focused on four "cornerstone" missions, Soho/Cluster, XMM-Newton, Herschel and Rosetta, and several smaller missions including Planck, Mars-Express, Huygens, Integral, ISO, Hipparcos \citep{inventing2017}. ESA budget for Earth observation increased consistently from 2008 to 2025 from 500 M\$(2023) to 2700 M\$ (2023), due to the investment of the European Commission in the Copernicus program. 

The cost of a standard science mission is driven by:
\begin{enumerate}
\item
{\bf Pushing scientific requirements to extremes} with little regard for engineering realities.  In absence of other guidelines scientists will always go to the maximum they can squeeze into a mission opportunity. This situation is a natural outcome of standard linear development approaches: first you consolidate the scientific requirements, then you produce a preliminary and followed by a final design able to satisfy those requirements, then you start the real production and finally you must demonstrate that what you have produced is compliant with the requirements set at the beginning. And for large missions the last step can follow many years, if not decades, after the first step. During which period both the original scientific requirements may have become obsolete, or they can be superseded by some technological innovation that has intervened in the meantime. 

The traditional development scheme is convenient for scientists, because once they have defined the objectives and scientific requirements, they don't then have to defend them day by day, and problems that may arise in developing a mission are deferred to engineers. This normally implies delays and cost growth (in addition to an engineer’s headache), both of which cannot be tolerated in the new space economics scheme. The attitude of scientists is not driven by laziness. They always want a better measurement. Given the long development time of missions and their consequent extreme rarity it is natural that they are not willing to compromise too much for that mission. Scientists know that any descoping will result in a net loss of science because there will be no opportunity to develop a successor mission in a reasonable time scale.
\item
{\bf Minimizing  launch mass and volume} because of high launch costs and limited fairing volumes. This need produces more advanced and complex engineering solutions,   JWST's "origami" unfolding is the prime example - if a "no-fold" mirror had been possible when JWST was designed no one would have chosen to fold it and thereby introduce hundreds of single point failure modes. 
\item
{\bf Use of space qualified parts.} As an example, the RAD750 computer that operates in many NASA missions including the Perseverance Mars rover costs several hundred thousands dollars, and is based on 1990 technology with a PowerPC 130-200MHz processor, $\sim$260-400 DMIPS\footnote{RAD750® radiation-hardened PowerPC microprocessor BAE Systems \url{https://www.baesystems.com/en-media/uploadFile/20210404045936/1434555668211.pdf}}. As a comparison, the computer used by the Ingenuity helicopter on board Perseverance is based  on the Qualcomm Flight board, which uses a smartphone computer chip, the Qualcomm Snapdragon 801 processor with a 2.5GHz clock, some 20 times more powerful and 20 times cheaper than the RAD750\footnote{\url{https://www.qualcomm.com/news/onq/2021/03/journey-mars-how-our-collaboration-jet-propulsion-laboratory-fostered-innovation}}. The Snapdragon 801 worked flawlessly for the about five years of cruise and operations of Ingenuity.
\item 
{\bf Failure is not an Option} leads to large numbers of quality control procedures to reduce the risk of failure. As an example, the ECSS (European Cooperation for Space Standardization) manuals commonly used by ESA and other European space agencies foresee that $>15000$ requirements for standard missions (e.g. Euclid) have to be met, mostly on quality standards. This is an expensive and time consuming procedure that greatly reduces the probability of  failures, but does not eliminate them. This is demonstrated by the problems that several ESA missions have had to deal with, the most recent ones being on Euclid\footnote{\url{https://www.space.com/euclid-guidance-light-sensor-setbacks-commissioning-dark-universe}
\url{https://europeanspaceflight.com/esa-euclid-telescope-has-an-ice-problem/}} and BepiColombo\footnote{\url{https://spacenews.com/esa-delays-bepicolombo-orbital-insertion-because-of-thruster-problem/}}. 
\item
{\bf Quasi-monopoly Suppliers.} Only a few companies win contracts for major science missions, effectively introducing a monopoly, which inflates costs. As an example, main prime contractors of ESA missions alternate between two companies (Thales-Alenia Space and Airbus) and major funding states (France, Germany, Italy and the UK, in the framework of the ESA geographical return principle). For a comprehensive report on the EU competitive gap in space see chapter 8 of the 2024 Draghi report\footnote{Draghi, M. The future of European competitiveness 2024 \url{https://commission.europa.eu/document/download/ec1409c1-d4b4-4882-8bdd-3519f86bbb92_en?filename=The\%20future\%20of\%20European\%20competitiveness\_\%20In-depth\%20analysis\%20and\%20recommendations_0.pdf}}.
\end{enumerate}

\subsection{Metrics for science mission productivity and impact}

\begin{figure}[ht]
\includegraphics[width=7.5cm]{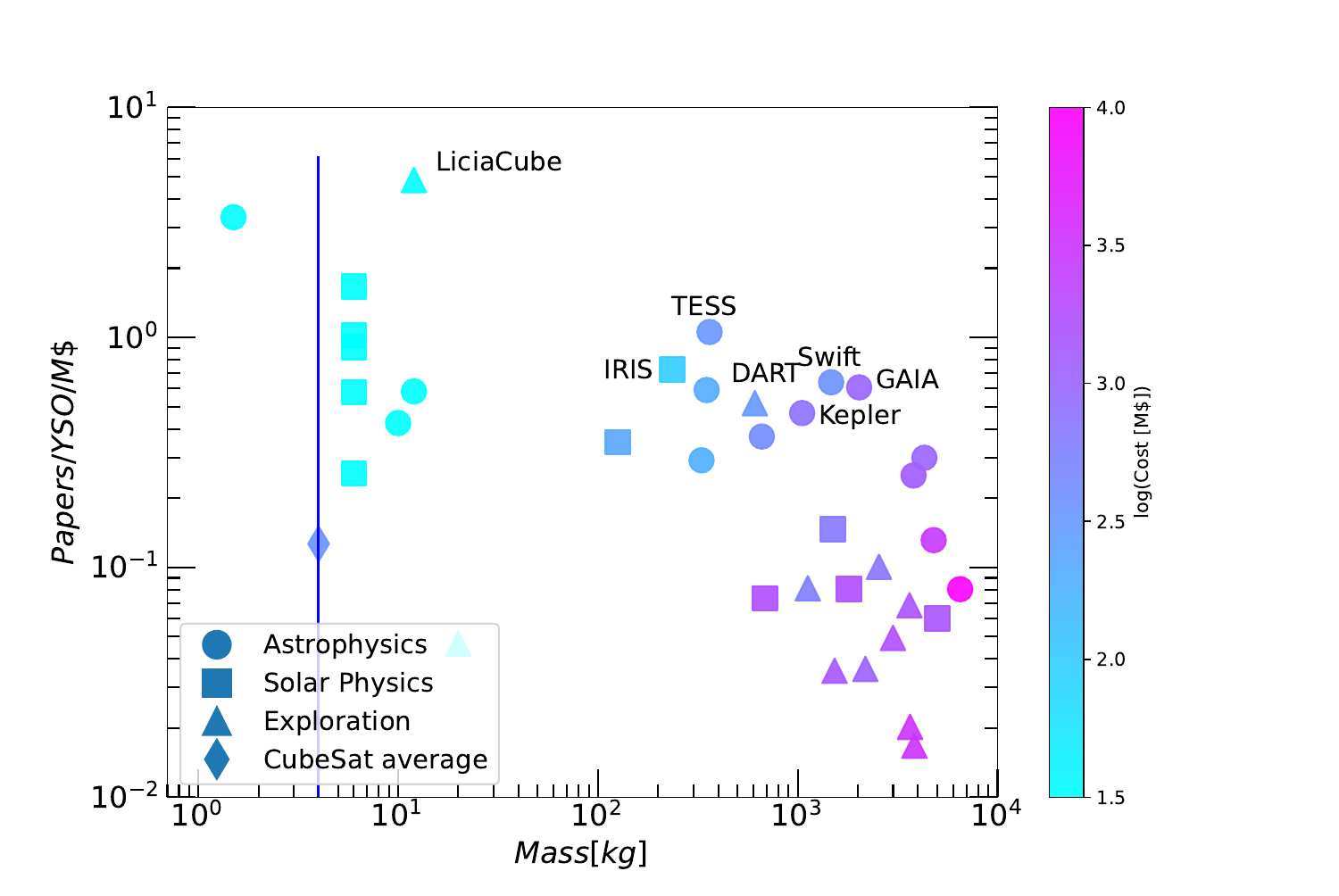}
\includegraphics[width=7.5cm]{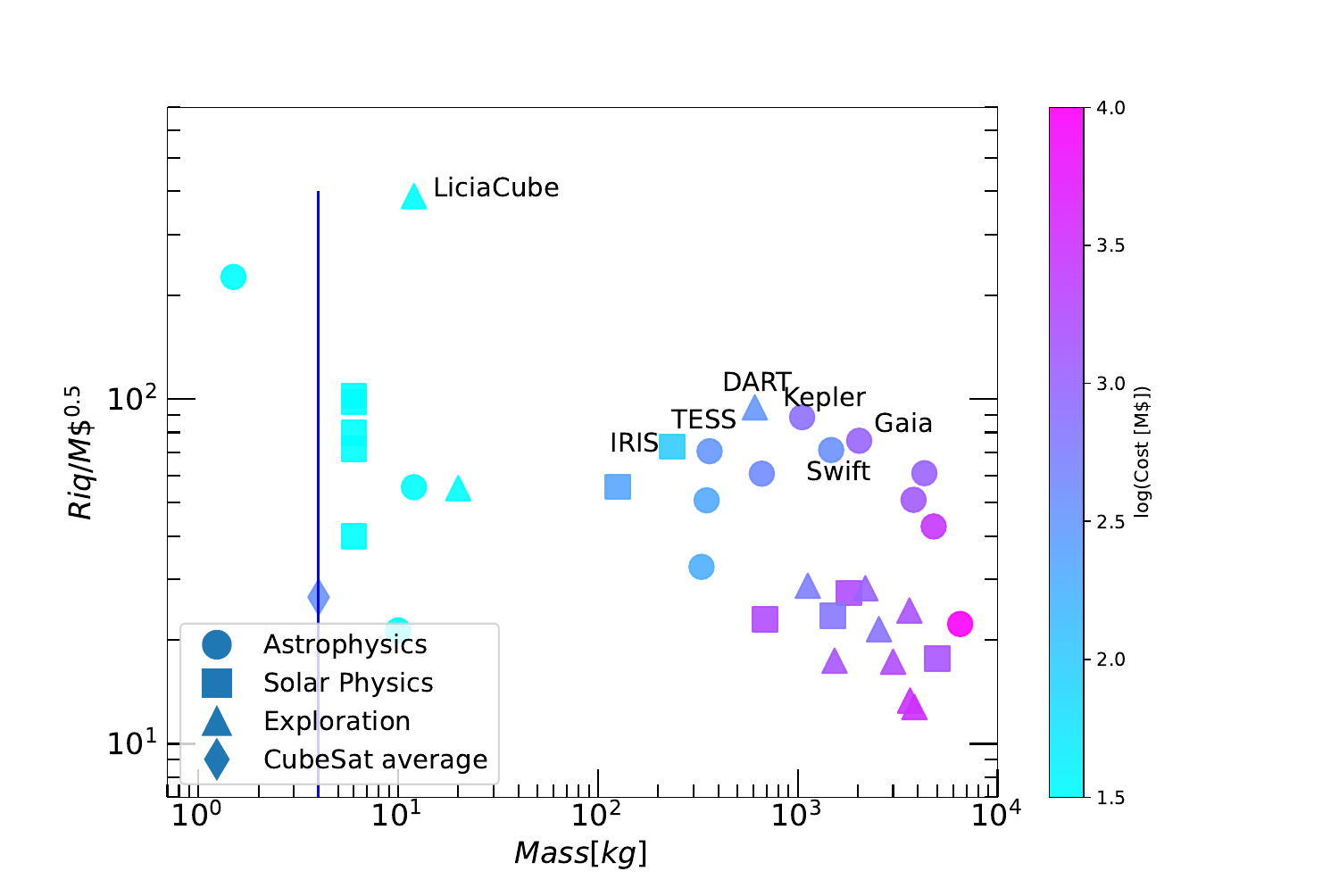}
\caption{[left panel] Productivity of space science missions as a function of their mass and cost (color). Productivity is parameterized as the number of reviewed papers/yr/M\$. The diamond marks the average of cubesat missions, the error bars encompass the range of productivity of cubesats, from very high to very low or zero for failed missions. [Right panel] Impact of space science missions as a function of their mass and total cost (color). Impact is parameterized as the Research impact quotient (Riq), see Appendix A. The diamond marks the average of cubesat missions, the error bars encompass the range of impact of cubesats, from very high to very low or zero for failed missions.
}
\label{sciencemissions}
\end{figure}

The high cost of standard space missions should be balanced by a high scientific return. It is not easy to provide an unbiased and quantitative measure of the scientific return of a mission. A standard metric for scientific productivity is the number of refereed papers produced based on a given mission, while a standard metric for evaluating the impact is the number of citations these papers have received. These metrics have the advantage that they are relatively easy to obtain but have important limitations; see Appendix A for a description on data collection and a discussion of their value. Figure \ref{sciencemissions} shows the productivity and impact of several recent space missions as a function of their mass and cost.  Productivity is parameterized as the number of articles per year since operation (YSO) per M\$ of total cost (bus, payload, launch), the impact is parameterized as the Research Impact quotient (Riq) divided by the square root of the total cost in M\$, see Appendix A for definitions.

Missions score similarly in the two metrics. By these measures, small/medium science missions, mainly NASA Explorer missions, tend to have better productivity and impact per M\$ than large flagship missions. The larger number of papers and citations of flagship missions does not fully compensate for their much higher cost. Among missions with costs greater than about 1 B\$, Gaia stands out for both productivity and impact, as also mentioned in Section 1 above. 

Cubesats can reach both top and bottom productivity and impact. They can be incredibly successful but also a complete failure, and they represent well the high-risk/high-gain concept behind the new space economy approach, emphasizing the value of a production line. 

Of course, simple metrics cannot tell the full story. Large, ambitious missions are more likely to produce transformational results than small, limited missions. Therefore, the real problem is whether space science can take advantage of the benefits of the new space economy also for medium/large missions in addition to nano/micro/small missions. In particular, can ambitious science missions exploit the three pillars of new space?
\begin{enumerate}
\item technology innovation. For example using greater mass, volume to orbit (at no increase in launch cost) to lower payload cost; efficiency in  innovation;
\item
business innovation. Efficiency in work organization and procurement, for example flight proven COTS and EEE components, vertical integration, and scale production;
\item
cultural innovation, with constrained scientific requirements: go for the good and not for the best; iterate requirements between scientists and  engineers; less demanding quality control procedures. If scientists can be convinced that relaxing requirements will lead to a flight mission much sooner, then they may be persuaded to take the 'bird in the hand'.
\end{enumerate}

\section{Toward a new generation of ambitious and affordable space science missions}

The development of several cubesat scientific projects in the past few years is a first attempt to take advantage of some of the benefits of the new space economy. The cubesat concept was proposed in 1999 by R. J. Twiggs at Stanford University. The original idea was to provide an educational framework for developing a satellite mission. This implies a short development time, consistent with a master's or Ph.D. degree courses, and a low cost, affordable by a university. The main limitation at that time was the launch cost per kg, which drove the requirement of having spacecraft of just a few kilograms.  After more than 20 years of development, cubesats have moved out of the purely educational sphere to be used for both commercial, scientific, and even defense applications. The total cubesat mission cost is in many cases small in absolute value, and thus affordable to Universities and research institutes teams, without the need of large investment by traditional space agencies. As an example, the total cost of the HERMES Pathfinder project, including flight hardware and software development, launch, operation, and science is $\sim$12MEUR, approximately a fourth paid from a European Commission Horizon 2020 grant, and the rest paid from an Italian Ministry of University and Research grant, and from the Agenzia Spaziale Italiana (ASI) funds. 

The relatively low cost of scientific cubesat projects has produced a proliferation of initiatives, which definitely increased the community working on space science, both in Academia and in Small, Medium Enterprises (SME) (for example, \cite{spence2022} and \cite{Bloser2022} present reviews of cubesat missions in Solar Physics and High Energy Astrophysics). The cost reduction of current cubesat projects is mainly the result of four factors:
\begin{itemize}
\item
miniaturization (less mass, smaller volumes);
\item
standardization of hardware components, using COTS components, use of flight proven EEE instead of radhard EEE. Radiation tests are sometimes used on most critical components or subsystems;
\item
reduced cost of manpower in Universities and research institutes with respect to space industries. Dedicated contracts include PhD and Post-Doc fellowships, while senior staff FTE are in many cases provided as in kind contributions by institutes and Universities; 
\item
less demanding quality control procedures. As an example, in the ECSS framework, 10 times less requirements are foreseen for ESA IOD cubesats. Furthermore, ECSS is often used as a reference only in cubesat contracts. Higher risks are allowed even within the ESA system for cubesats.
\end{itemize}

However, while having high impact/dollar (Section 3), scientific cubesats are still rather expensive in terms of cost/kg (section 2). In particular, the cost of cubesat missions with ambitious requirements like HERMES Pathfinder. The high cost/kg of HERMES Pathfinder has three main contributing factors:
\begin{itemize}
\item
ambitious scientific and system requirements: temporal resolution of 300-400ns, 5-7 times better than that of Fermi/GBM; extremely wide energy band from a few keV to a few MeV; sensitivity similar to that of Fermi/GBM; S/C 3 axis stabilized, ADCS system from an higher category (6U); UHF/VHF, S-band and Iridium communications; $>$two year lifetime;
\item
use of COTS components at relatively expensive market price;
\item
limited cost saving for development of many flight models.
\end{itemize}

In contrast, Camelot pathfinder missions (GRBAlpha, VZLUSAT2, GRBBeta see\cite{Pal2020, Ripa2022}), have been developed following an iterative approach, delivering at each stage a minimum viable product, but reaching orbit in a very short time and very low price, still producing relevant scientific results \cite{Pal2023, Ripa2023}. 

So far, the developers of scientific cubesat projects have not been able to take advantage of two key elements of cost reduction in the commercial space economy: scale production and vertical integration. The cost of COTS components procured on the market, although much smaller than space-qualified components, remains relatively high. As launch costs continue to decline (Section 2.2) the limitations of mass and volume will be surmountable. The cubesat approach could then be applied to larger spacecraft. We are already witnessing this trend: the number of small satellites predicted to launch in the next decade is declining as typical satellites get heavier\footnote{\url{https://spacenews.com/heavier-smallsats-weigh-down-market-forecasts/}}. The first possible candidates for scientific applications are the smallsats from 50kg to 300kg, that enter the range of the NASA Explorer missions. So far, the realization of the Explorer mission S/C has been contracted to traditional space companies\footnote{Ball Aerospace and technology  for IXPE, SPHEREx;  Nortrop Grumman or Orbital ATK for NuStar, TESS, ICON, COSI; Lockheed Martin for MUSE, respectively}. Today, new space companies are exploiting scale production (multi-purpose buses) and vertical integration to offer satellite buses in the 100-1000kg range at low prices. Examples are K2 Space\footnote{\url{https://www.k2space.com}} a company with the goal of providing large satellite buses (P/L up to 1 ton) at low prices ($<$\$15M) and short procurement time (months instead of years or even decades); and APEX\footnote{ \url{https://www.apexspace.com/}}, with the aim of providing buses to host payloads in the 100-150 kg range at prices of 4-8M \$ and with a procurement time of half a year. Exploiting such standardized buses, without major customization, can make scientific missions cheaper and much more frequent. 

Space-based observatories and exploration probes that are developed on timescales of years instead of decades can respond much more quickly to emerging scientific problems. More frequent space missions can revolutionize the way space science is done because they can trigger a virtuous loop between opening up new fields of investigation and finding timely responses to the new questions. 

There are two main conditions that must be satisfied to enable this paradigm shift. 

{\bf (1) a wider funding base}. In the old space economy, nearly all basic science missions were paid by governments, and in particular by space agencies, NASA and ESA first among them. And space agencies have their settled ways of working on science missions (see Table 5), which will take time to change, although elements of change have been seen both in the past (the faster, better, cheaper mantra of then NASA Administrator Dan Goldin during the 1990s [21], and in the present (as in the already mentioned case of the rethink of MSR and VIPER). 

In contrast,the new space economy is mainly driven by private funding. Private investors usually look for commercial returns on short time scales, while the return from basic science is definitely on a long time scale, if any. So, a sudden shift from public to private funding seems unlikely. What scientists can do is to work toward diversifying funding from other entities such as:
\begin{enumerate}
\item 
private investors, interested in advertising their activities or because they might see profit opportunities in activities such as planetary protection (Space Weather Events and asteroids), or exploration of Moon, Mars, and asteroids for resource prospecting, or using science mission to test novel sensors and advanced technologies in orbit to be then applied to commercial activities. Scientists could reap real benefits from their ability to generate useful knowledge for the private sector;
\item 
philanthropic foundations.  In the past private foundations have supported large ground-based telescopes, with costs in the range of tens and even hundreds million \$ \cite{MacDonald2017}. With similar, or even less, money today it is foreseeable to develop powerful scientific space missions;
\item 
public-private partnership (PPP), with private companies providing advanced technological capabilities as services. PPP was the basis for innovation in space infrastructure, starting with the NASA Commercial Orbital Transportation Services (COTS) program in the 2000s \cite{Anderson2023, garver2022}. Other examples of very fruitful PPPs in the past are PNT and Earth Observation, where public funding started or helped develop the systems, and private funding helped develop the market \cite{Anderson2023}. 
\item
inter-governmental organizations like the European Commission. The last EC work plan that included Space in the 'Horizon Europe' program is focused mainly on four items: scientific return from GNC (Galileo); scientific return from Earth observation (Copernicus); development of new satellite communication systems (GOVSATCOM); and development of quantum technologies\footnote{\url{https://ec.europa.eu/info/funding-tenders/opportunities/docs/2021-2027/horizon/wp-call/2023-2024/wp-7-digital-industry-and-space_horizon-2023-2024\_en.pdf}}. 
\end{enumerate}
In the first three cases, technological development starts with TRL 5 or 6. The development of new technologies (TRL 0-5) would remain funded through government funds and, at least partially, from intergovernmental organizations, although philanthropy may contribute. This is of course a key issue because transformational results are often obtained using novel technologies which had to be invented on purpose.

{\bf (2) Innovation:} in the new space economy innovation must proceed to ensure the largest possible return in the shortest possible time. This is made possible by both business and cultural innovation. Culturally, it is crucial to:
\begin{enumerate}
    \item 
make the requirements smart. If we want to develop in a systematic way basic science missions in the new space context, we must understand that our scientific requirements must not be carved in stone. Where you end is not where you start. Constraints and requirements can become obsolete, surpassed, and therefore no longer relevant. Sometimes the lesson that 'we did it this way and it worked, so we should do the same again' is not the right lesson for new circumstances. While choosing ambitious science objectives, we must at the same time push for the good, the solid, the robust rather than the best imaginable. Cost should be a major factor in choosing our approach. Once we are able to achieve the good in a short timescale, we can improve it in the next step. Of course, descoping is a concept well known to the scientific community and has been applied to many missions in recent decades. Here, though, it is not simply a matter of recognizing at some point that a particular science or system requirement is not achievable at the expected cost or time and eliminating or simplifying it. Rather, it is about planning from the beginning an iterative process to achieve the main science goal, during which the science requirements are continually challenged and questioned, with rapid iteration at each step of the development process. This is quite a radical change with respect to the way a science mission is developed today at NASA and ESA, and to the ways most scientists working on space missions are used to. However, if missions become much more frequent and with much shorter development times, requirements compromises make a lot of sense to speed up the entire process; we can more than recover the science loss in the compromise in a new mission. 
\item 
Agree that scientific and system requirements must be continuously challenged during the development of a project opens the way to optimize processes.  Before looking for complex technical solutions, we must be sure that they are really necessary to reach our main scientific objectives. If not, we can save a lot of time and money simply by not tackling those complex technical problems. Iterative development saves time and money and at the same time makes us more confident about the identified solutions.
\item 
Once all technical solutions have been identified, we can proceed to speed up development by reducing procedures and automating the process. 
\item
Many academic institutions are not designed to be able to collaborate efficiently with the commercial world. There is a lack of adequate information on business logic in many cases, and a lack of a true risk-oriented culture. We often fall into two extremes: most of the times risk is only a virtual concept that remains on paper; some time however we tend to put all our eggs in one basket. In any case risk and failure are not usually considered positive attributes, and we rarely follow the approach of learning from mistakes. Furthermore, most academic institutions must deal with rules for the recruitment of personnel and for procurement that are much more cumbersome and stringent than for private companies. Therefore, acceleration of scientific project development times probably requires the setting up of spin-offs or public-private organizations. 
\end{enumerate}

The above points imply a major paradigm shift. This shift is unlikely to be made quickly and comprehensively.  It is probably more efficient to identify case studies first. The large programs in Moon exploration of both Western nations and China and its allies, with the Artemis and ILRS programs, offer a path forward. Scientific projects associated with these lunar programs could be placed as a first-case test bed for the changes proposed here. Major resources have been brought into play for lunar programs by both governments and private investors in the hope of creating a truly new lunar space economy. Indeed, according to a study by the Center for Strategic and International Studies, the size of the lunar economy during the next decade will be of the order of 8-10B\$/yr\footnote{\url{https://csis-website-prod.s3.amazonaws.com/s3fs-public/2024-10/241021_Swope_Swimmimg_Upstream_0.pdf}}. 

Scientists can 'sell' the knowledge needed to harvest resources on the Moon, leveraging the ability of the same knowledge to have a high scientific value. In other cases the instruments used for lunar prospecting can be re-tuned to respond to other ambitious scientific questions. Definitely, a lot of good science can be done on or from the Moon \citep{Crawford2012,Silk2023}, and it would be truly bizarre if this science did not benefit from the emerging lunar space economy. Lunar science and lunar economy can also be good examples of full public-private partnership \cite{Sommariva2020}.

\section{Conclusions}

In the past 10 years, in the same time interval that JWST was launched and new ambitious space science missions like the Habitable World Observatory and the Mars Sample Return mission were conceived, the so-called 'new space' economy, driven by private funding, has emerged powerfully. Today, the new space economy has surpassed the government-funded space economy. The introduction of a market logic in space activities has resulted in a dramatic reduction in cost and schedule. 

The new space economy was stimulated by public-private partnerships during the 2000s, thanks to new initiatives by NASA, where they experimented with radical changes in direction, such as buying services from private companies with fixed-price contracts, instead of developing programs led by NASA itself and contracted out to industry for execution with cost-plus contracts \cite{Anderson2023, garver2022}. In recent years, the same approach was followed by the US Department of Defense. As an example, the National Reconnaissance Office is taking advantage of the SpaceX Starlink assembly line to build a network of spy satellites on a very short timescale and at a low price, the Starshield project\footnote{\url{https://arstechnica.com/space/2024/04/spacex-working-with-northrop-grumman-on-spy-satellites-for-us-government/}}.
Private companies like K2 Space exploit vertical integration and scale production to offer huge '3-meter-by-3-meter' payload bay at less than \$ 15M per satellite, with a lead time of a few months \footnote{\url{https://techcrunch.com/2024/12/19/k2-space-will-fly-its-extra-large-satellite-for-the-first-time-in-2026/}}. 

Scientists might profitably exploit this paradigm shift to realize ambitious missions at a cost 10 times or more smaller than that of today's standard large NASA and ESA missions. Working on two main issues can foster this paradigm shift: 

(1) broaden the funding base. In addition to government money, scientists can fund their projects through philanthropic foundations, public-private partnerships, and private investors. Knowledge is a crucial space resource, and it is precisely the resource that basic science researchers are best at harvesting. Accurate Position, Navigation, and Timing, the largest business in today's and also tomorrow's space economy, moving hundreds billion dollars, is famously impossible without general relativity corrections. Ironically, Albert Einsten (the inventor of general relativity more than 100 years ago) and his heirs did not make a single dollar from this invention. More relevant to new space is the story of Riccardo Giacconi, 2002 Nobel Prize winner for the discovery of Cosmic X-ray sources and the Cosmic X-ray Background. When Giacconi joined Bruno Rossi of MIT in 1959 as a young researcher, he was hired by American Science and Engineering (AS\&E), a company started the year before by Martin Annis and of which Bruno Rossi was chair of the board of directors. AS\&E was working on NASA contracts, but was also very active in the market. They developed the first body scanner, later adopted in airports around the world. Money earned in the market allowed AS\&E to let Rossi and Giacconi develop the strategies and instrumentation that led to the creation of X-ray astronomy and to a series of world-leading missions: Uhuru, Einstein Observatory, and Chandra \citep{Giacconi2008}. This is one of the several possible scenarios that could link science and business.  Today, the new space economy probably provides more opportunities than in the time of Rossi and Giacconi. Two broad areas seem particularly promising today and merit a mention: quantum technologies and resource prospecting on the Moon, Mars, and asteroids. 

(2) the ability of scientists to keep up with the innovations of the new space economy. Technology innovation proceeds through alternating phases of incremental and disruptive development. Business innovates through vertical integration and scale production, but also a service-oriented business model and a culture of innovation that manifest itself as openness to risk and iterative development. Giacconi again comes to our aid in highlighting these concepts. Soon after the discovery of a strong, seemingly uniform, X-ray background in 1962 Giacconi understood that to detect the sources responsible for the Cosmic X-ray background, an imaging quality of the order of an arcsecond was needed in X-rays, several thousands times better than then available. At those early times, it was difficult, if not impossible, to convince NASA to embark directly on a project to develop and launch an X-ray observatory with arcsecond capabilities. Instead, Giacconi imagined an iterative program aimed at consolidating the field and arriving at the arcsecond telescope (Chandra) through steps, Uhuru first, to provide the first all-sky survey of bright X-ray sources at the beginning of the 1970s, Einstein Observatory, to demonstrate in space the capabilities of X-ray optics, and finally the Advanced X-ray Astrophysics Facility, AXAF, renamed Chandra after launch in 1999. The new space economy offers the opportunity to perform iterative approach programs of this kind on much shorter timescales and at much lower cost. 

\appendix
\section{}
\label{sec:appendix}
We have searched the Astrophysical Data System (ADS) and the SCOPUS database for refereed publications, citations, and the H index of a sample of recent space scientific missions funded by NASA and ESA. We included in the list a sample of cubesats realized by Universities and scientific institutes. The search was carried out using the simplest possible keywords, to be able to have a data sample as homogeneous as possible. As an example, the Gaia ADS Library produced by ESA includes 12.842 refereed papers since 2014. The library is obtained by merging the results of a complex and customized ADS query with a manually curated library. The authors claim a fraction of false positives $<$5\%. The much simpler ADS query ((facility:Gaia OR abs:Gaia OR title:Gaia) year:2014-2024 property:refereed collection:astronomy) produces 6617 entries, of which $>$90\% are present in the ESA Gaia ADS Library. A manual check on the papers missing in the ESA Gaia ADS Library shows that a large fraction are actually using Gaia data or results obtained from Gaia data. So we are confident that the fraction of false positives in our table is certainly smaller than 10\% and probably smaller than 5\%. Limiting the sample to papers that explicitly mention the name of the mission in the title, abstract, or facility keyword would include most papers that are based on or make particularly important use of data or results from that specific mission. Papers not included are likely to be those that make indirect use of the data from that particular mission or use the data only in an ancillary way. We have chosen to use queries that on the one hand ensure the inclusion in the samples of the majority of papers that make the greatest use of data from each mission, and on the other hand that are as simple as possible to ensure the greatest homogeneity when comparing samples from different missions. 

The fraction of false positives is a strong function of the mission name. Searching for missions with familiar names such as “Fermi” or “Swift”, or “Dice” produces a large number of false positives. To limit them as much as possible, we added search keywords relevant to the missions. For example, in the search for “Fermi” we required the presence in the full text of the name of at least one of the Fermi instruments or targets: (facility:Fermi) OR (abs:Fermi AND (full:GRB OR full:"Fermi LAT" OR full:"FERMI/LAT" OR full:"Fermi GBM" OR full:"Fermi/GBM")) year:2008-2024 property:refereed collection:astronomy. We stopped tailoring the query when the results converge to within a few \%. Finally, we manually inspected the final paper list to look for obvious false positives.

We verify that the results obtained from the ADS database are consistent with those obtained from the SCOPUS database. We find that the variations are always smaller than 20\%. 

We use the ratio of the number of refereed articles to the number of years of operation, normalized for the total cost of missions in M\$ to parameterize mission productivity. 

Following \cite{Pepe2012} we use the Research impact quotient $Riq=\sqrt{tori}/YSO$ to parameterize the impact of a mission. $YSO$ is again is again the number of years from operation. $tori$ is the total impact of the research given by $tori=\sum_{cit} 1/ar$, where $cit$ is the collection of citations, $a$ is the number of authors in each article cited and $r$ is the number of references in the article cited. The $tori$ is defined as the amount of work devoted to exploitation of a given mission calculated in research papers. To assess the impact of the mission normalized for the mission cost, we use $Riq/\sqrt{M\$}$.

\noindent
{\bf Acknowledgments}
This work was performed in part at the Aspen Center for Physics, which is supported by the National Science Foundation grant PHY-2210452. FF acknowledges support from the INAF grant PRORIS. FF acknowledges useful discussions with Norbert Werner, Salvo Sciortino, Paolo Tozzi, Francesca Esposito. 

 \bibliographystyle{elsarticle-num} 
 \bibliography{cas-refs}





\end{document}